\documentclass[prd,preprint,superscriptaddress,preprintnumbers,nofootinbib]{revtex4}
\usepackage{graphicx}
\usepackage{epsfig}
\usepackage{bm}
\usepackage{latexsym,amssymb,amsmath,amssymb,wasysym,float}
\usepackage{mathrsfs}  
\usepackage{color}

\usepackage{enumitem}

\newcommand{\postscript}[2]{\setlength{\epsfxsize}{#2\hsize}
   \centerline{\epsfbox{#1}}}

\usepackage[usenames,dvipsnames]{xcolor}
\definecolor{orange}{cmyk}{0,0.5,1,0}
\definecolor{rossoCP3}{cmyk}{0,.88,.77,.40}
\definecolor{graa}{rgb}{0.8,0.8,0.8}
\definecolor{blaa}{rgb}{0.2,0.2,0.6}

\def\Ra{=\!\!\!>}

\def\simlt{\mathrel{\lower2.5pt\vbox{\lineskip=0pt\baselineskip=0pt
           \hbox{$<$}\hbox{$\sim$}}}}
\def\simgt{\mathrel{\lower2.5pt\vbox{\lineskip=0pt\baselineskip=0pt
           \hbox{$>$}\hbox{$\sim$}}}}

\newcommand{\sbt}{\,\begin{picture}(-1,1)(-1,-3)\circle*{1.45}\end{picture}\ }     
\newcommand{\be}{\begin{equation}}
	\newcommand{\ee}{\end{equation}}
\newcommand{\ba}{\begin{eqnarray}}
	\newcommand{\ea}{\end{eqnarray}}

 \allowdisplaybreaks

\begin{document}

\title{\color{rossoCP3} Large extra dimensions from higher-dimensional inflation}

\author{\bf Luis A. Anchordoqui}

\affiliation{Department of Physics and Astronomy,  Lehman College, City University of
  New York, NY 10468, USA}

\affiliation{Department of Physics,
 Graduate Center, City University
  of New York,  NY 10016, USA}

\affiliation{Department of Astrophysics,
 American Museum of Natural History, NY
 10024, USA}

\author{\bf Ignatios Antoniadis\footnote{Present address: Simons Center for Geometry and Physics,
Stony Brook University, Stony Brook NY 11794, USA}
}

\affiliation{School of Natural Sciences, Institute for Advanced Study,\\ Princeton, NJ 08540, USA}

\affiliation{Laboratoire de Physique Th\'eorique et Hautes \'Energies - LPTHE,\\
Sorbonne Universit\'e, CNRS, 4 Place Jussieu, 75005 Paris, France}




\begin{abstract}
\noindent 
We propose the possibility that compact extra dimensions can obtain large size by higher dimensional inflation, relating the weakness of the actual gravitational force to the size of the observable universe. Solution to the horizon problem implies that the fundamental scale of gravity is smaller than $10^{13}$ GeV which can be realised in a braneworld framework for any number of extra dimensions. However, requirement of (approximate) flat power spectrum of primordial density fluctuations consistent with present observations makes this simple proposal possible only for one extra dimension at around the micron scale. After the end of five-dimensional inflation, the radion modulus can be stabilised at a vacuum with positive energy of the order of the present dark energy scale. An attractive possibility is based on the contribution to the Casimir energy of right-handed neutrinos with a mass at a similar scale.
\end{abstract}

\maketitle

\newpage
\section{Introduction}
In this work, we study the possibility of connecting different large hierarchies appearing in particle physics and cosmology, which are expected to be described by the same fundamental theory of Nature. On the one hand, cosmological observations suggest that the observable universe must have undergone a period of rapid expansion in order to make compatible its size with causal connection and explain the horizon and homogeneity problems. On the other hand, in particle physics, a major hierarchy is related to the actual weakness of gravitational relative to gauge interactions. Another hierarchy is related to the smallness of neutrino masses which appear to be of the same order of magnitude as the dark energy scale.

A way to connect these hierarchies between particle physics and cosmology is via the size of extra dimensions which are necessary ingredients for consistency of string theory. Indeed, if their size is large compared to the fundamental (string) length, the strength of gravitational interactions becomes strong at distances larger than the actual four-dimensional (4D) Planck length~\cite{Arkani-Hamed:1998jmv, Antoniadis:1998ig}. As a result, the string scale is detached from the Planck mass consistently with all experimental bounds if the observable universe is localised in the large compact space~\cite{Antoniadis:1998ig}.

The possibility that large extra dimensions may be related to the smallness (in Planck units) of some physical scales~\cite{Antoniadis:1988jn} is also supported in the context of the swampland program~\cite{Vafa:2005ui} by the distance/duality conjecture, stating that large distances in the string landscape of vacua imply a tower of exponentially light states, such as the Kaluza-Klein (KK) tower associated to the decompactification of extra dimensions~\cite{Ooguri:2006in}. The underlying assumption is that these physical scales are determined by vacuum expectation values of moduli fields with an exponential dependence in the proper distance. The tower of states provides a 4D effective field theory (EFT) description up to the so-called species scale $M_*$, where gravity becomes strong and an ultraviolet completion of the theory is needed~\cite{Dvali:2007hz} (such as string theory).

Our proposal is that compact dimensions may have undergone a uniform rapid expansion, together with the three-dimensional non-compact space, by regular exponential inflation driven by an (approximate) higher dimensional cosmological constant~\cite{Anchordoqui:2022svl}. Their size has grown from the fundamental length to a much larger value, so that at the end of inflation the emergent 4D strength of gravity became much weaker, as it is measured today. 
We show that this idea implies that the higher dimensional gravity scale $M_*$ is less than $10^{13}$ GeV but above 10 TeV for any number of extra dimensions $1\le d\le 6$, consistently with accelerator constraints on new physics, laboratory tests of Newton's law and astrophysical constraints (which impose higher lower bounds on $M_*$ for $d=1$ and $2$)~\cite{ParticleDataGroup:2022pth}.

A period of higher-dimensional inflation can be easily realised by introducing a bulk $d$-dimensional inflaton with an appropriate scalar potential having a sufficiently flat region, as one usually considers in four dimensions. Besides solving the horizon problem, 4D slow-roll inflation predicts an approximate scale invariant power spectrum of primordial density perturbations consistently with  observations of the cosmic microwave background (CMB). This is due to the fact that the 2-point function of a massless minimally coupled scalar field in de Sitter (dS) space behaves logarithmically at distances larger than the cosmological horizon, a property which is though valid for any spacetime dimensionality~\cite{Ratra:1984yq}. When some dimensions are however compact, this behaviour is expected to hold for distances smaller than the compactification length, while deviating from scale invariance at larger distances, potentially conflicting with observations at large angles. One would therefore expect that a period of 4D inflation is necessary for generating the required scale invariant power spectrum of primordial perturbations.

Here, we show that consistency with CMB observations is maintained if the size of extra dimensions is larger than about a micron, implying a change of behaviour in the power spectrum at angles larger than 10 degrees, corresponding to multiple moments $l\simlt 30$, where experimental errors are getting large~\cite{Planck:2018vyg}. This amazing numerical coincidence singles out the case of $d=1$ extra dimension of micron size, with a corresponding species scale $M_*\sim 10^9~{\rm GeV}$, as the only possibility for the simplest realisation of our proposal consistent with observations. 

An interesting question is to understand the above result from the 4D perspective. It turns out that uniform 5D exponential expansion corresponds in 4D Planck units to a power law inflation where the scale factor of the non-compact 3D space expands as $t^3$ in terms of the 4D Friedman-Robertson-Walker (FRW) metric proper time $t$, while the radius of the extra dimension expands as $t^2$. This implies that within the 5D theory, one needs less number of e-folds of rapid cosmological expansion in order to solve the apparent 4D horizon problem. Moreover, scale invariance of the power spectrum~\cite{Harrison:1969fb, Zeldovich:1972zz} is obtained upon summation over the contribution of the inflaton KK-modes' fluctuations that correspond to a tower of scalars from the 4D point of view.

This scenario can be nicely combined with the Dark Dimension proposal for the cosmological constant using the distance/duality conjecture within the swampland program~\cite{Montero:2022prj}. It can also realise an old idea for explaining the smallness of neutrino masses by introducing the right-handed neutrinos as 5D bulk states with Yukawa couplings to the left-handed lepton and Higgs doublets that are localised states on the Standard Model brane stack~\cite{Dienes:1998sb, Arkani-Hamed:1998wuz, Dvali:1999cn}; the neutrino masses are then suppressed due to the wave function of the bulk states.

After the end of 5D inflation, the radion modulus has a runaway exponential quintessence-like potential  with an exponent near the upper allowed value for accounting the present dark energy, assuming a local minimum of the 5D scalar potential along the inflaton direction with positive energy of order $10^{-2}$ eV. Alternatively, the radion can be stabilised in a local (metastable) dS vacuum, using the contributions of bulk field gradients~\cite{Arkani-Hamed:1999lsd} or of the Casimir energy, assuming a mass for the bulk R-handed neutrinos of the same order of magnitude~\cite{Arkani-Hamed:2007ryu}. Such a bulk mass was considered in the past in order to weaken the bounds on the size of the extra dimension from an analysis of the neutrino oscillation data~\cite{Lukas:2000wn,Lukas:2000rg,Carena:2017qhd,Anchordoqui:2023wkm}.

\section{Higher-dimensional inflation and its 4D interpretation}
We start with $(4+d)$-dimensional gravity having $d$ compact extra dimensions of size $R$:
\be
S_{4+d}=\int[d^4x\, d^dy] \left({1\over 2}M_*^{2+d}{\cal R}^{(4+d)}-\Lambda_{4+d}\right)
\label{Shd}
\ee
where ${\cal R}^{(4+d)}$ is the higher dimensional curvature scalar with $M_*$ the corresponding reduced Planck mass and $\Lambda_{4+d}$ a positive higher dimensional cosmological constant. For notational simplicity, we use brackets in the measure transforming as a density under 5D diffeomorphisms. Upon dimensional reduction to four dimensions, the effective 4D metric in the Einstein frame is obtained by the line element decomposition
\be
ds^2_{4+d} = \left({r\over R}\right)^{d} ds_4^2 + (R/R_0)^2 ds_d^2\,,
\label{metric}
\ee
where the internal volume associated to $ds_d^2$ is normalised to $(2\pi R_0)^d$, with $R_0$ the initial value of the radius at some time instance $\tau=\tau_0$, $r\equiv\langle R\rangle$ after the end of inflation, while we neglected the graviphotons and kept only the overall internal volume $(2\pi R)^d$.\footnote{Here, we also consider the internal space to be locally flat, which is automatic for $d=1$}. 
The resulting 4D action is:
\be
S_{4}=\int[d^4x] \left({1\over 2}M_p^{2}{\cal R}^{(4)}-{d(d+2)\over 4}M_p^2\left({\partial R\over R}\right)^2
-(2\pi r)^{2d}{\Lambda_{4+d}\over (2\pi R)^{d}}\right)\,, 
\label{S4d}
\ee
with the 4D reduced Planck mass $M_p$ and the scalar potential $V$ 
defined by
\be
M_p^2=M_*^{2+d}(2\pi r)^d \quad;\quad V={M_p^2\over M_*^{2+d}}{\Lambda_{4+d}\over (R/r)^{d}}
\label{MpV}
\ee

The vacuum solution of the higher dimensional theory \eqref{Shd} is a $(4+d)$-dimensional de Sitter spacetime, describing in flat slicing a uniform exponentially expanding universe of external and internal dimensions. The corresponding line element in conformal coordinates is conformally flat:
\be
ds^2_{4+d} = {\hat a}^2(\tau)\, d{\hat s}^2_{4+d} \quad;\quad {\hat a}(\tau)={1\over H\tau}\,,
\ee
where $d{\hat s}_{4+d}$ stands for the flat $(4+d)$-dimensional line element and ${\hat a}(\tau)$ is the conformal factor depending on the conformal time $\tau$ and the Hubble parameter $H$ given by ${(3+d)(2+d)\over 2}H^2=\Lambda_{4+d}/M_*^{2+d}$. Using the metric decomposition \eqref{metric}, one can express the uniform $(4+d)$-dimensional inflation into a time-dependent 4D metric in the Einstein frame  $ds^2_4=a^2(\tau)\, d{\hat s}^2_4$ and a corresponding time-dependent background for the radius of the internal dimensions $R(\tau)$:
\be
ds^2_{4+d} ={a^2(\tau)\over (R(\tau)/R_0)^d}\,d{\hat s}^2_4 + \left({R(\tau)\over R_0}\right)^2\, ds^2_d\,,
\label{lineelement}
\ee
implying 
\be
a(\tau)=\left(R(\tau)/R_0\right)^{1+d/2} = ({\hat a}(\tau))^{1+d/2}
\quad;\quad R(\tau)={\hat a}(\tau)R_0=R_0\left(a(\tau)\right)^{2/(2+d)}\,,
\label{aRexp}
\ee
where the normalisation of the conformal factor was fixed by the initial condition $a(\tau_0)={\hat a}(\tau_0)=1$ at $\tau_0=H^{-1}$.
Comparing with~\eqref{metric}, this 
implies that at the beginning of inflation the universe was of the size of the higher-dimensional fundamental length $M_*^{-1}$ which is $(r/R_0)^{d/2}$ larger than the actual 4D Planck length.


In FRW coordinates, where the higher-dimensional cosmic time is ${\hat t}=-H^{-1}\ln(H\tau)$, both the 3D scale factor of the 4D universe and the radius of the internal space expand exponentially: 
\be
{\hat a}({\hat t})=e^{H{\hat t}} \quad\Ra\quad a({\hat t})= 
e^{(1+d/2)H{\hat t}} \quad;\quad R({\hat t})=e^{H{\hat t}}R_0\,.
\ee
As a result, $N$ e-folds of 4D expansion emerge from $2N/(d+2)$ e-folds of the $(4+d)$-dimensional metric leading to $2N/(d+2)$ e-folds of the internal radius expansion. 

Note however that the 4D cosmic time $t$ is different from $\hat t$ since $a(\tau) = (H\tau)^{-(1+d/2)}$. It follows that
\ba
\!\!\!\!\!\!\!\!\!
Ht= {2\over d}
(H\tau)^{-d/2}\quad&\Ra&\quad H{\hat t}={2\over d}\ln \left({d\over 2}
Ht\right) \,,
\nonumber\\[10pt]
&&\hskip 0.5cm  a(t)=\left({d\over 2} 
Ht\right)^{1+2/d} \quad;\quad R(t)=R_0\left({d\over 2} 
Ht\right)^{2/d} \,,
\label{higherDbg}
\ea
which gives for example:
\ba
\label{5Dbg}
d=1&:& a(t)= \left({1\over 2}
Ht\right)^3\quad;\quad R(t)= R_0 \left({1\over 2} 
Ht\right)^2 \,,\\[5pt]
d=2&:& a(t)=(Ht)^2\quad;\quad R(t)= R_0 Ht \, .\nonumber
\ea
It follows that uniform exponential expansion in higher dimensions corresponds to a power-law inflation from the 4D perspective, together with an expanding size of the compact dimensions parametrised by a time-dependent radion modulus.

Combining Eqs.~\eqref{MpV} and \eqref{aRexp}, one has:
\be
M_*=M_p (2\pi r M_*)^{-d/2}\quad;\quad r=R_0 a^{2/(2+d)} = R_0 e^{2N/(2+d)} \,,
\label{MpMstar}
\ee
where $N$ is the number of e-folds of expansion of the 3D space at the end of the higher dimensional inflation. It follows that
\be
M_*=M_p e^{-dN/(2+d)}\,.
\ee
Solution to the horizon problem requires that $N\simgt 30-60$ when inflation occurs above the TeV and below $M_p$, implying $M_*\simlt 10^{13}$ GeV.\footnote{An approximate formula is $N\simgt \ln{M_I\over{\rm eV}}$ where $M_I$ is the inflation scale.} On the other hand, imposing $M_*\ge 10$ TeV, one obtains
\be
14\,(\ln10)\ge{d\over d+2} N\quad\Ra\quad N\simlt 32\left(1+{2\over d}\right) \,,
\ee
which can be easily satisfied for any $d$. However in the special cases of $d=1,2$, one obtains stronger bounds on $M_*$ from non accelerator experiments. Indeed, for $d=1$ laboratory tests of Newton's law imply $r\simlt 30\,\mu$m  corresponding to $M_*\simgt 10^8~{\rm GeV}$~\cite{Lee:2020zjt}, while for $d=2$ astrophysics constraints from supernovae imply $M_*\simgt 10^6$ GeV corresponding to a radius $r\simlt 0.1$ nanometer ($r^{-1}\simgt 10$ keV)~\cite{Hannestad:2003yd}.

From Eq.~\eqref{aRexp}, it follows that a co-moving distance $d(x,x')$ between two points $x$ and $x'$ in 4D space at equal times corresponds to a $(4+d)$-dimensional physical distance 
$d(x,x'){\hat a}(\tau)=d(x,x')a(\tau)^{2/(d+2)}=d(x,x')a(\tau)(R_0/R)^{d/2}$, leading to the relation:
\be
d^{\,\tau}_\text{phys}(x,x')=d(x,x')\,a(\tau)=d(x,x')\,{\hat a}(\tau)\left({R\over R_0}\right)^{d/2}
=\hat{d}^{\,\tau}_\text{phys}(x,x')\,
{M_p\over M_*}\,,
\label{distanceconversion}
\ee
where the extra factor in the relation between the  physical distances $d^{\,\tau}_\text{phys}$ and $\hat{d}^{\,\tau}_\text{phys}$, in 4 and $d+4$ dimensions, can be understood as the change of mass units from the 4D to the $(4+d)$-dimensional Planck length, $M_p^2(\tau) =(2\pi R(\tau))^d M_*^{2+d}$. 

We now turn to the discussion of the CMB power spectrum. Precision of observational data applies to angles of less than about 10 degrees, corresponding to the distance Mpc scale ($\sim\!\!10^{22}$~m), equivalent to Gpc distance today. On the other hand, the number of e-folds during radiation dominance is about 29 when the inflation scale $M_I$ is at TeV~\cite{efolds}, corresponding to an expansion of about $4\times 10^{12}$. The Mpc distance was then around $10^9$~m. Assuming that the reheating and radiation evolutions have the same scaling~\cite{Cook:2015vqa}, the above distance is scaled down  by a factor TeV$/M_I$ at a higher inflation scale, while converting it to higher-dimensional units using \eqref{distanceconversion}, it is scaled down by an additional factor $M_*/M_p$. Assuming $M_I\sim M_*$, one finds a scaling factor of TeV$/M_p\sim 10^{-15}$ leading to the distance scale of a micron which is miraculously of order of the size of the 5th dimension for $d=1$ at the end of inflation. Thus, compatibility with the CMB power spectrum is expected to work only for $R\simgt 1\,\mu$m which selects the case of one extra dimension $d=1$ as the only possibility that does not require a period of 4D inflation. In the following, we specialise our analysis to this case.\footnote{Note that for $d=1$ the metric decomposition~\eqref{metric} is exact when the extra dimension is compactified on an interval $S^1/Z_2$ since the graviphoton is projected out of the spectrum.} 

\section{Density perturbations from inflation in five dimensions}
During 5D inflation, the line element for spatially flat metric is:
\ba
ds_5^2&=&{\hat a}^2(\tau)\left[-d{\tau}^2+d\vec{x}^2+dy^2\right]\nonumber\\
&=&R_0{a^2(\tau)\over R(\tau)}\left(-d\tau^2+d\vec{x}^2\right)+(R(\tau)/R_0)^2dy^2\,,
\ea
with
\be
R(\tau)=R_0\,{\hat a}(\tau)\quad;\quad a(\tau)={\hat a}^{3/2}(\tau)\,.
\label{background5}
\ee


Density fluctuations can be computed from the equal-time 2-point function of a massless minimally coupled scalar field $\Phi$ in de Sitter spacetime which is given by (in Fourier space)~\cite{Ratra:1984yq}:
\be
\langle\Phi^2({\hat k},\tau)\rangle={\pi\tau\over 4\,{\hat a}^3}\left[ J_\nu^2({\hat k}\tau)+Y_\nu^2({\hat k}\tau)\right]\,,
\label{Phi2}
\ee
where $\hat k$ is the co-moving space momentum given by ${\hat k}^2=k^2+n^2/R_0^2$ with $k$ the 3-space momentum (wavenumber) and $n$ the KK number, while $J_\nu,Y_\nu$ are Bessel functions of order $\nu=(D-1)/2=2$ for $D=5$.
For small argument (large wave lengths compared to the dS Hubble radius $H^{-1}$), the Bessel functions behave as:
\ba
J_\nu(z)&\simeq& {1\over \Gamma(\nu+1)}\left({z\over 2}\right)^\nu\to 0\quad;\quad \nu>0 \,,\\
Y_\nu(z)&\simeq& -{1\over\pi}\left({2\over z}\right)^\nu
-{1\over\pi}\left({2\over z}\right)^\nu\sum_{l=1}^{\nu-1}{(\nu-l-1)!\over l!}{\left(z\over 2\right)}^{2l}+{2\over\pi}J_\nu(z)\ln{z\over 2}+{\cal O}(z^\nu)\quad;\quad \nu\in Z\nonumber\\
&\to&-{1\over\pi}\left({2\over z}\right)^2\left(1+{z^2\over 4}\right)\quad;\quad \nu=2 \, .
\ea
As a result
\be
\langle\Phi^2({\hat k},\tau)\rangle\simeq{\tau\over 4\pi\,{\hat a}^3}\left(1+{4\over {\hat k}^2\tau^2}\right)^2\sim
{4\over\pi}{H^3\over ({\hat k}^2)^2} \, .
\ee
The 2-point function at the Standard Model brane position, located for instance at the origin of the 5th dimension, is obtained by performing the sum over the KK excitations in ${\hat k}^2$:
\ba
\sum_n{1\over \left({n^2\over R_0^2}+k^2\right)^2}&=&-{\partial\over\partial k^2}\sum_n{1\over {n^2\over R_0^2}+k^2}
=-{\pi R_0\over 2k}{\partial\over\partial k}\left[{1\over k}\coth(\pi kR_0)\right]\nonumber\\
&=&{\pi R_0\over 2k^2}\left({1\over k}\coth(\pi kR_0)+{\pi R_0\over\sinh^2(\pi kR_0)}\right)\,,
\ea
which leads to:
\be
\langle\Phi^2(k,\tau)\rangle_{y=0}\simeq {2R_0 H^3\over k^2}
\left({1\over k}\coth(\pi kR_0)+{\pi R_0\over\sinh^2(\pi kR_0)}\right)\,.
\ee

In order to get an estimate of the argument $\pi kR_0$, one can express it in terms of the corresponding 5D physical wavelength $\hat\lambda=2\pi{\hat a}/k$ and use the fact that $R/\hat\lambda$ remains the same (time independent) in co-moving and physical 5D frames. Note however that the 4D physical wave length (in the 4D Einstein frame) is $\lambda=2\pi{a}/k=(R/R_0)^{1/2}\hat\lambda$ which is larger than $\hat\lambda$ by a factor $M_p/M_*$ and is time dependent. One can therefore evaluate $\pi kR_0$ at the end of inflation, in the 5D Einstein frame. It turns out that $\pi kR_0=2\pi^2R/\hat\lambda>1$ for any $\hat\lambda$ less than a micron when the size of the extra dimension is in the range $R\sim 1-10\,\mu$m. In this region the 4D physical wave lengths $\lambda$ are less than about a kilometre, as argued above. 

One can now see that the 2-point function changes behaviour for momenta $k$ around the compactification scale, as we expected. At large momenta (`small' wave lengths), it scales as $1/k^3$ since $\coth(\pi kR_0)\to 1$ and the resulting amplitude $\cal A$ of the power spectrum $\cal P$ is scale invariant:
\be
\pi k R_0>1 \quad\Ra\quad
{\cal A}={k^3
\over 2\pi^2}\langle\Phi^2(k,\tau)\rangle_{y=0}\simeq {R_0H^3\over\pi^2}\sim {H^2\over\pi^2}\,,
\label{P1}
\ee
where for the last estimate $R_0$ was chosen of order $H^{-1}$ for simplicity. 
The result is therefore the same as the one from the standard cosmology, but obtained from 5D inflation. Small violation of scale invariance can be introduced as usually, for instance by a small tachyonic mass (as in hilltop inflation), leading to a spectral index slightly below unity.

On the other hand, for very large wavelengths corresponding to physical distances in CMB much bigger than Mpc, the argument $(\pi k R_0)<1$ and the behaviour of the 2-point function changes to:
\be
\pi k R_0<1 \quad\Ra\quad
\langle\Phi^2(k,\tau)\rangle_{y=0}\simeq{4H^3\over\pi 
k^4}\quad;\quad {\cal A}\simeq{2H^3\over\pi^3 k}\,,
\label{P2}
\ee
leading to a vanishing spectral index (or to almost vanishing by deviating slightly from de Sitter space). 


\section{Interpretation of the results from a 4D perspective}

A higher dimensional inflaton corresponds, during inflation around a flat region of the scalar potential, to an approximately massless, minimally coupled scalar field in dS space, leading to the $(4+d)$-dimensional action: 
\be
S_{4+d}^{\rm inflaton}=\int[d^4x]\,d^dy\left[-{1\over 2}(\partial_\mu\Phi)^2 +{1\over 2}\left({R\over R_0}\right)^{-d-2}(\vec{\partial}_y\Phi)^2\right]\,,
\ee
where Eq.~\eqref{metric} was used and we computed explicitly the radius dependence, while kept the 4D metric associated to the line element $ds_4^2$ as arbitrary background in the measure and for contracting the 4D indices.
This action should be considered in the 4D background \eqref{aRexp} with the normalisation change~\eqref{lineelement} corresponding to a higher dimensional dS spacetime, for instance \eqref{5Dbg} in the 5D case. Thus, in FRW conformal coordinate frame, it reads in Fourier space:
\ba
S_{4+d}^{\rm inflaton}&=&\int d\tau\, a^2(\tau)\, d^3k\sum_{\{n\}}\left\{{1\over 2}\,
\dot{\tilde \Phi}_n^2(k) -{1\over 2}\left[ k^2
+a^2(\tau)\left({R_0\over R}\right)^d{n^2\over R^2}\right]{\tilde \Phi}_n^2(k)\right\}\nonumber\\
&=&\int d\tau\, a^2(\tau)\, d^3k\sum_{\{n\}}\left\{{1\over 2}\,\dot{\tilde \Phi}_n^2(k) -{1\over 2}\left(k^2
+ {n^2\over R_0^2}\right){\tilde \Phi}_n^2(k)\right\}\,,
\ea
where dot stands for derivative with respect to the conformal time $\tau$, $\{n\}$ indicates summation over all components of the vector, and we dropped the vector symbol to simplify notation. 
Note that expressing the co-moving product $kR_0$ in terms of physical quantities, one has for $d=1$:
\be
kR_0={k\over a}({\hat a}R_0){a\over{\hat a}}=2\pi{r\over\lambda}\left({r\over R_0}\right)^{1/2}\,,
\ee
where the last equality holds at the end of inflation with $r$ the size of the extra dimension at the micron scale. The factor $(r/R_0)^{1/2}$ scales the physical distances at the end of 5D inflation 
to 4D units, as pointed out in~\eqref{distanceconversion}.

We now define $\Phi=\chi/a$ and thus $\dot\Phi=(\dot\chi-({\dot a}/a)\chi)/a=(\dot\chi+3\chi/(2\tau))/a$ since $a\sim\tau^{-3/2}$ from \eqref{aRexp}. It follows that
\ba\label{inflatonKK}
S_{5}^{\rm inflaton}=\int d\tau\, d^3k\sum_n\left\{{1\over 2}\,\dot{\tilde \chi}_n^2(k)+{1\over 2}\left[
{15\over 4}{1\over\tau^2}-\left(k^2 + {n^2\over R_0^2}\right){\tilde \chi}_n^2(k)\right]
\right\}\,,
\ea
which is precisely the action of a massless minimally coupled scalar in 5D de Sitter spacetime~\cite{Ratra:1984yq}, leading to the solution~\eqref{Phi2}. We stress again that the product $kR_0\ne pR$ for physical 4D momenta $p=k/a$ since $a\ne \hat a$,
as explained above.

Note that reducing \eqref{inflatonKK} to the 0-mode contribution $n=0$ does not lead to a scale invariant power spectrum. Indeed, the coefficient 15/4 differs from the 4D value of 2 for a massless field around dS$_4$ and corresponds to a 4D metric background of power law inflation $a(t)\sim t^3$, as in \eqref{5Dbg}, or equivalently by an inflaton around dS$_4$ near a maximum of its potential with tachyonic mass $\sqrt{7}H/2$; for details, see Appendix. The resulting spectral index vanishes and
is already excluded by CMB observations~\cite{Planck:2013jfk}. Scale invariance is restored only upon summation over all KK modes corresponding to a tower of 4D `inflatons'. This summation is implied for fluctuations of fixed 3D wave length observed at the position of the brane. 

\section{Large-angle CMB power spectrum}

We now turn to estimate the CMB power
spectrum at large angular scales. We begin by noting that a general operator ${\cal O}_{w}$ with conformal weight $w$ has an equal time correlation function, 
\begin{equation}
\langle{\cal O}_{w} (\vec x) {\cal O}_{w} (\vec x')\rangle
\sim | \vec x-\vec x'|^{-2w} \,,
\end{equation}
and a power spectrum in Fourier space given by
\begin{equation}
\tilde G_2 (|\vec k|; w) = \langle\tilde{\cal O}_{w} (\vec k) \tilde {\cal O}_{w}(-\vec k)\rangle \sim 
\int d^3 \vec x \  e^{i \vec k\cdot \vec x}\,|\vec x|^{-2w}\,
\sim  |\vec k|^{2w - 3} \, .
\end{equation}
The spectral index of the energy density fluctuations in Fourier space
is $n_s = 2 w - 3$, and so writing as usual ${\cal P} \propto {\cal A}_s (k/k_*)^{n_s -1}$, from (\ref{P1}) and (\ref{P2}) it
follows that $n_s = 1$ for the scale-invariant Harrison-Zel'dovich
spectrum~\cite{Harrison:1969fb,Zeldovich:1972zz} and $n_s = 0$ for the
5D de Sitter spectrum at large angular scales, yielding $w = 2$ and
$w= 3/2$, respectively. Here ${\cal A}_s$ is the variance of curvature
perturbations in a logarithmic wavenumber interval centered around the
pivot scale $k_*$. For these large angular scales, the intrinsic
temperature fluctuations of the photons at decoupling are subdominant
with respect to the gravitational potential perturbation they encounter
{\it en route} to Earth, and so the scaling weight of the perturbations in
the gravitational potential grows into $s = w-2$~\cite{Sachs:1967er}.

\begin{figure}[tpb]
\postscript{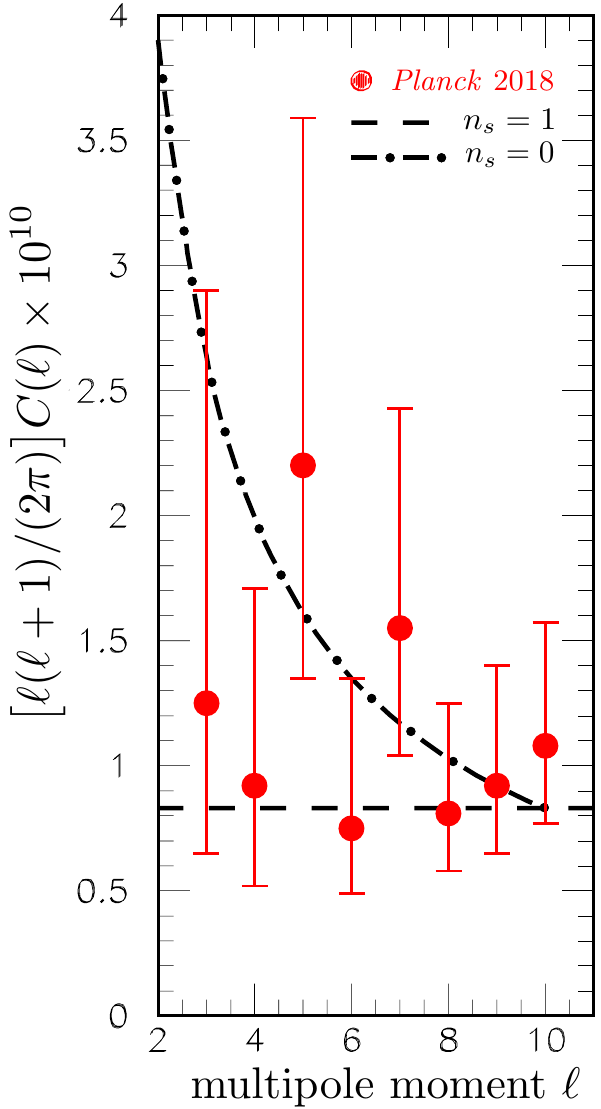}{0.5}
\caption{Large-scale CMB power spectrum as reported by the Planck
  Collaboration~\cite{Planck:2018vyg}, superimposed over the predictions for $n_s =1$ and
  $n_s =0$. The $n_s=1$ curve has been normalized to the {\tt Plik} best
  fit to Planck 2018 data, i.e. $\ln (10^{10} {\cal A}_s) =
  3.0448$~\cite{Planck:2018vyg}, whereas the $n_s=0$ curve has been normalized to match the $n_s =1$
  curve at $\ell =10$. \label{fig:1}}
\end{figure}

Assuming that the statistical distributions of matter and metric
fluctuations around the background metric are isotropic, the two-point
correlation function (for directions $\hat n$ and $\hat n'$)  of CMB
temperature  fluctuations $\delta T/T$ is
driven by the scaling dimension $s$, 
\begin{eqnarray}
G_2^{\rm CMB}(\hat n\cdot\hat n'; s) &\equiv &\left\langle\frac{\delta T}{T}(\hat n)
\frac{\delta T}{T}(\hat  n')\right\rangle
\sim \int d^3 \vec k\left(\frac{1}{|\vec k|^2}\right)^2 \tilde G_2 (|\vec k|; s + 2) e^{i \vec k\cdot 
(\vec x- \vec x')} \nonumber \\
&=& c_{s} \Gamma (-s) (1 - \hat n \cdot\hat n')^{- s}\,,
\label{c}
\end{eqnarray}
where $c_s$ is an $s$-dependent constant~\cite{Antoniadis:2011ib}. The
multipole expansion for general $s$ gives \begin{equation}
G_2^{\rm CMB}(\hat n \cdot \hat n'; s) = \frac{1}{4\pi} \sum_{\ell =1}^{\infty} (2\ell + 1)
C_s (\ell) P_{\ell} (\hat n \cdot \hat n')\ ,
\end{equation}
where
\begin{equation}
C_s({\ell}) \sim \Gamma(-s) \sin\left(\pi s\right)
\frac{\Gamma (\ell + s)}{\Gamma (\ell + 2 - s)}\ ,
\end{equation}
with a pole singularity at $s =0$ appearing in the $\ell = 0$ 
monopole moment. However, CMB anisotropy maps are constructed by
removing the monopole and the dipole contributions, and so the $\ell
=0$ term does not appear 
in the multipole series expansion. The higher moments of the anisotropic two-point 
correlation are well-defined for $s \to 0$. Normalizing to the quadrupole moment
$C_s(2)$, we find
\begin{equation}
C_s ({\ell}) = C_s(2)\,
\frac{\Gamma (4 - s)} {\Gamma (2+ s)} \,
\frac{\Gamma (\ell + s)}{\Gamma(\ell + 2 - s)}\,  .
\label{Cell}
\end{equation}
Eq.~(\ref{Cell}) provides a reasonable
approximation of the
anisotropy power spectrum for $\ell \lesssim
30$, and for $n_{_s}=1$, it reduces to
the textbook example~\cite{Dodelson:2003ft}
\begin{equation}
C ({\ell}) = 6 C (2) \ \frac{1}{\ell (\ell +1)} = \frac{2 \pi}{25} {\cal
  A}_s \ \frac{1}{\ell (\ell +1)} \, .
\end{equation}
For larger $\ell$ multipoles, an exhaustive transfer-function analysis (numerically solving the
linearized Einstein-Boltzmann equations) would be required. However,
herein we are interested in distances $\gg {\rm Mpc}$, which
corresponds to $\ell \lesssim 10$. In Fig.~\ref{fig:1} we show a
comparison of the predictions for $n_s =1$ and $n_s =0$, which are
confronted to the latest data from the {\it Planck} 
mission~\cite{Planck:2018vyg}.

\section{End of inflation}

In order to stop the exponential expansion in higher dimensions, one may introduce an inflaton $\Phi(x,y)$ as a bulk field with a potential having a typical slow-roll region~\cite{Anchordoqui:2022svl}. Assuming some mechanism of radius stabilisation, at the end of inflation the universe should be populated by appropriate decays of the inflaton into brane (Standard Model) fields that should dominate its decay to gravitons. For this, one may introduce a direct coupling of $\Phi$ to brane fields localised at the brane position, such as Yukawa couplings to fermions, etc. By dimensional analysis, the corresponding decay rate is
\be
\Gamma^\Phi_{\rm SM}\sim {y^2\over (RM_*)^d} m_\Phi\,,
\ee
where $m_\Phi$ is the inflaton mass, $y$ denotes the (dimensionless) coupling of the inflaton to the brane, while the denominator corresponds to the volume suppression. On the other hand, the decay rate of $\Phi$ to gravitons occurs gravitationally in $(4+d)$ dimensions 
with corresponding decay rate
\be
\Gamma^\Phi_{\rm grav}\sim {m_\Phi^{3+d}\over M_*^{2+d}}\,.
\ee
Requiring the gravitational decay to be less that $\Gamma^\Phi_{\rm SM}$, one finds 
\ba
m_\Phi &<& \left( M_*^2/R^d\right)^{1/(2+d)}= 
M_*(M_*/M_p)^{2/(2+d)}\nonumber\\
&\simlt&\,{\cal O}(1)\,{\rm TeV}, \quad\text{for}\,\, d=1\,\, \text{and}\,\, M_*\sim 10^9\,\text{GeV} \, .
\ea
This model represents a specific realization of the dynamical dark matter framework~\cite{Dienes:2011ja}
in which the cosmic evolution of the hidden sector is primarily dominated by ``dark-to-dark'' decays with a small violation of the KK momentum conservation~\cite{Gonzalo:2022jac,Law-Smith:2023czn}. Alternatively, we can advocate a large violation of the KK momentum as in the fuzzy dark matter scenario discussed in~\cite{Anchordoqui:2023tln}.

\section{Radion stabilisation at the end of inflation $\bm{(d=1)}$}

At the end of inflation, the radion acquires a runaway potential emerging from the 5D cosmological constant $\Lambda_5^{\rm min}$ at the minimum of the inflaton potential: 
\be
V_0=2\pi r^2{\Lambda_5^{\rm min}\over R}\,,
\label{V0}
\ee 
where the line element decomposition \eqref{metric} was taken into account, see \eqref{S4d}.
Actually, $\Lambda_5^{\rm min}$ is constrained by the Higuchi bound~\cite{Higuchi:1986py} on the mass of the lightest spin-2 KK mode at the minimum of the potential:
\be
{1\over r^2}\ge 2H^2_{\rm min}={2\over 3}{\Lambda_4\over M_p^2}={4\over 3}\pi r{\Lambda_5^{\rm min}\over M_p^2}
\quad\Ra\quad \Lambda_5^{\rm min}\le {3\over 4\pi}{M_p^2\over r^3}\quad;\quad
(\Lambda_5^{\rm min})^{1/5}\simlt 100\,{\rm GeV}\,.
\label{V0bound}
\ee
$V_0$ has an exponential quintessence-like form, expressed in terms of the canonically normalised field in Planck units $\phi=\sqrt{3/2}\ln (R/r)$, see \eqref{S4d}; it is proportional to $e^{-\alpha\phi}$ with an exponent $\alpha\simeq 0.82$, which is curiously just at the upper limit of the experimentally allowed value~\cite{Barreiro:1999zs}. 

On the other hand, the radion can be stabilised by taking into account some generic additional contributions to its potential:
\be
V=\left({r\over R}\right)^2 {\hat V} + V_C \quad;\quad {\hat V}=2\pi R \Lambda_5^{\rm min} + T_4 + 2\pi{K\over R}\,,
\label{potential}
\ee
where 
the second contribution arises from localised 3-branes/orientifolds with total tension $T_4$, 
the third contribution arises from kinetic gradients of bulk 5D fields~\cite{Arkani-Hamed:1999lsd},\footnote{Actually, $K = \tilde K R_0^2 \sim \tilde K/M_*^2$ with $\tilde K$ of dimension five, so that $K$ has dimension three.} such as a scalar background linear in the fifth $(y)$ coordinate, while the last contribution $V_C$ corresponds to the Casimir energy from all 5D states of mass $m$~\cite{Arkani-Hamed:2007ryu} (with periodic boundary conditions):
\be
V_C=2\pi R\left({r\over R}\right)^2{\rm Tr}(-)^F\rho(R,m)\quad;\quad
\rho(R,m)=-\sum_{n=1}^\infty {2m^5\over (2\pi)^{5/2}} {K_{5/2}(2\pi Rmn)\over (2\pi Rmn)^{5/2}}\,,
\label{VC}
\ee
where $F$ denotes the fermion number and trace stands for the sum over all bosonic and fermionic 5D degrees of freedom. 

Since the Casimir potential falls off exponentially at large $R$ compared to the particle wave length while at small $R$ behaves as $1/R^4$, for simplicity, below we discuss two extreme cases of stabilisation: (i) neglecting $V_C$ and (ii) taking it into account for $K=0$.

\subsection{Stabilisation neglecting $V_C$}

By inspection of the expression \eqref{potential} neglecting $V_C$, one finds that a minimum of $V$ with (approximately) vanishing energy requires ${\hat V}'={\hat V}=0$ at the minimum, leading to
\be
r=\left({K\over\Lambda_5^{\rm min}}\right)^{1/2}\quad;\quad T_4=-4\pi(K\Lambda_5^{\rm min})^{1/2}
\quad;\quad V''={\hat V}''\mid_{R=r}=4\pi{K\over r^3}\,.
\label{minimum}
\ee
It follows that the tension must be negative and the radion mass is
\be
m_\phi^2={4\over 9}{r^2\over M_p^2} {\hat V}''\mid_{R=r}={4\over 9}{|T_4|\over M_p^2}
={8\over 9}{1\over r^2}{K\over M_*^3}\,.
\label{radionmass}
\ee
It is straightforward to compute also the maximum corresponding to
\be
R_{\rm max}=3r\quad;\quad V_{\rm max}={2\over 27}|T_4|\,.
\label{maximum}
\ee
Requiring the existence of a minimum with any vacuum energy less than $V_{\rm max}$ satisfying the Higuchi bound $m_{\rm spin-2}^2\ge 2H^2=2V_{\rm min}/(3M_p^2)$, it follows from \eqref{radionmass} and \eqref{maximum} that $m_\phi\le 3/r$. Thus, consistency with experimental bounds on extra forces implying a radion mass heavier than $0.1$ eV requires $K\sim M_*^3$ and ($\Lambda_5^{\rm min})^{1/5}\sim (K/r^2)^{1/5}\sim 100$ GeV, near its upper bound~\eqref{V0bound}. Moreover, from \eqref{minimum} one obtains $|T_4|^{1/4}\sim 1$ TeV so that all three terms of the potential \eqref{potential} are of the same order with tuneable vacuum energy at the minimum. 

Note that the presence of $V_C$, which is non-vanishing at least for the 5D graviton, does not affect this minimum, since its contribution is negligible compared to the others. However, for a different choice of parameters, such as in the case of vanishing $K$ that we discuss below, its contribution will be important for the minimisation of the potential. In fact, for values of $K$ much lower than $M_*^3$, the radion mass drops below the compactification scale $1/r$ and experimental limits on extra forces require a suppression of its coupling to matter that we also discuss in the next subsection.

\subsection{Stabilisation with $K=0$}

Neglecting $V_C$ at large $R$, the potential \eqref{potential} for $K$ vanishing has a maximum at 
\be
R_{\rm max}=-{T_4\over\pi\Lambda_5^{\rm min}}\,,
\label{Rmax}
\ee
requiring again a negative tension $T_4$. 
A minimum near the maximum can be then generated from the Casimir energy contribution along the lines of \cite{Arkani-Hamed:2007ryu}. Indeed $\rho(R,m)$ vanishes exponentially for $mR>1$, while for $mR$ small behaves as:
\be
\rho(R)\simeq -{1\over (2\pi R)^5\pi^2}\left[ \zeta(5) -{2\over 3}\pi^2\zeta(3) (mR)^2 + {\cal O}\left((mR)^4\right)\right]
\simeq -{1\over (2\pi R)^5\pi^2}\,,
\ee
where $\zeta(5)\simeq 1.037$, leading to
\be
V_C\simeq {2\pi r^2\over 32\pi^7R^6}(N_F-N_B)\,,
\ee
and where $N_F-N_B$ is the difference between the number of light fermionic and bosonic degrees of freedom which must be positive for a minimum to exist. Note from \eqref{5Dbg} that during inflation the canonically normalised field $\log R$ varies logarithmically with the 4D proper time and its speed decreases as $\sim 1/t$. It is therefore expected to oscillate around the minimum.

\begin{figure}[htb!]
  \begin{minipage}[t]{0.48\textwidth}
    \postscript{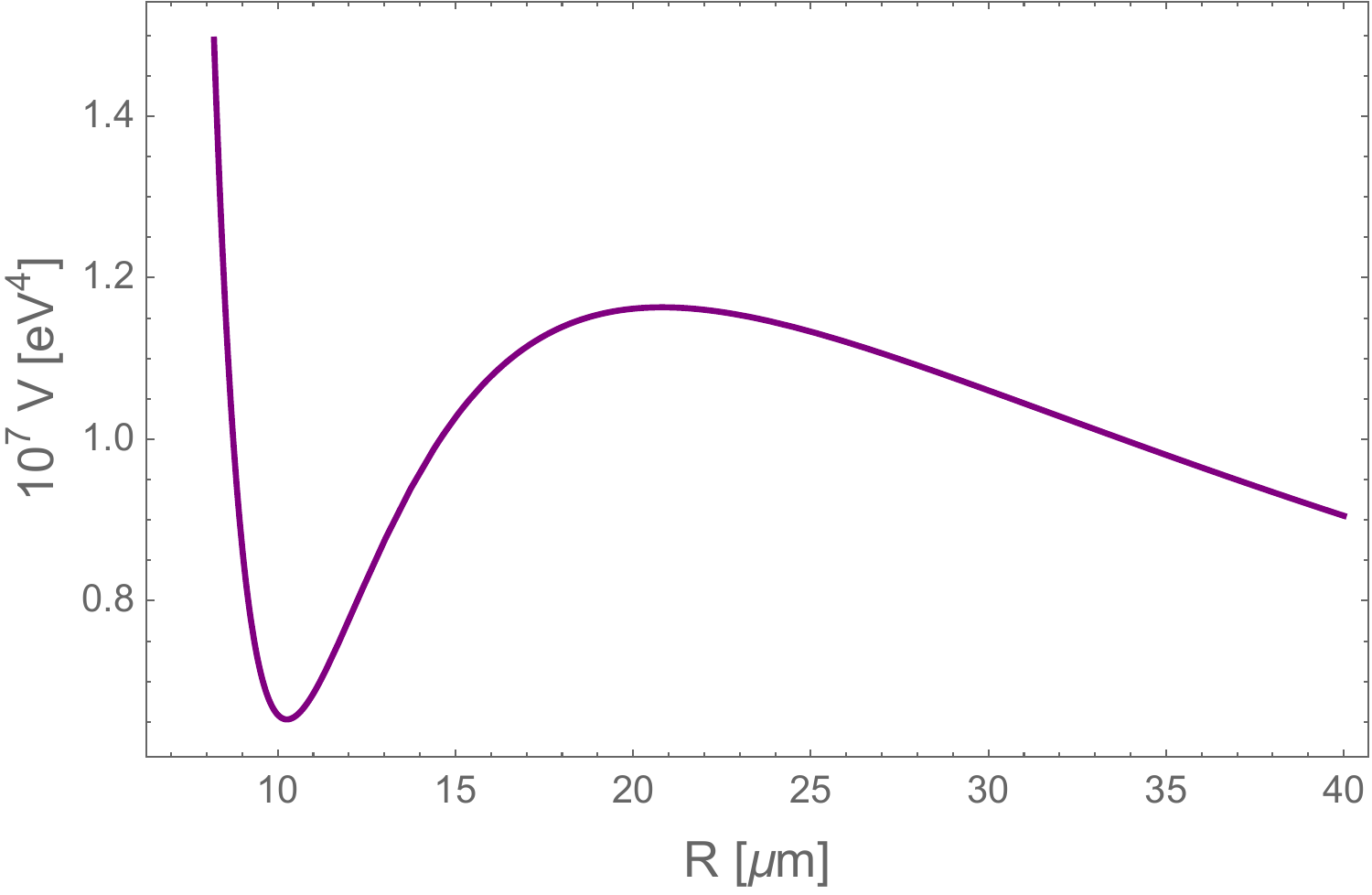}{0.8}
  \end{minipage}
\begin{minipage}[t]{0.48\textwidth}
    \postscript{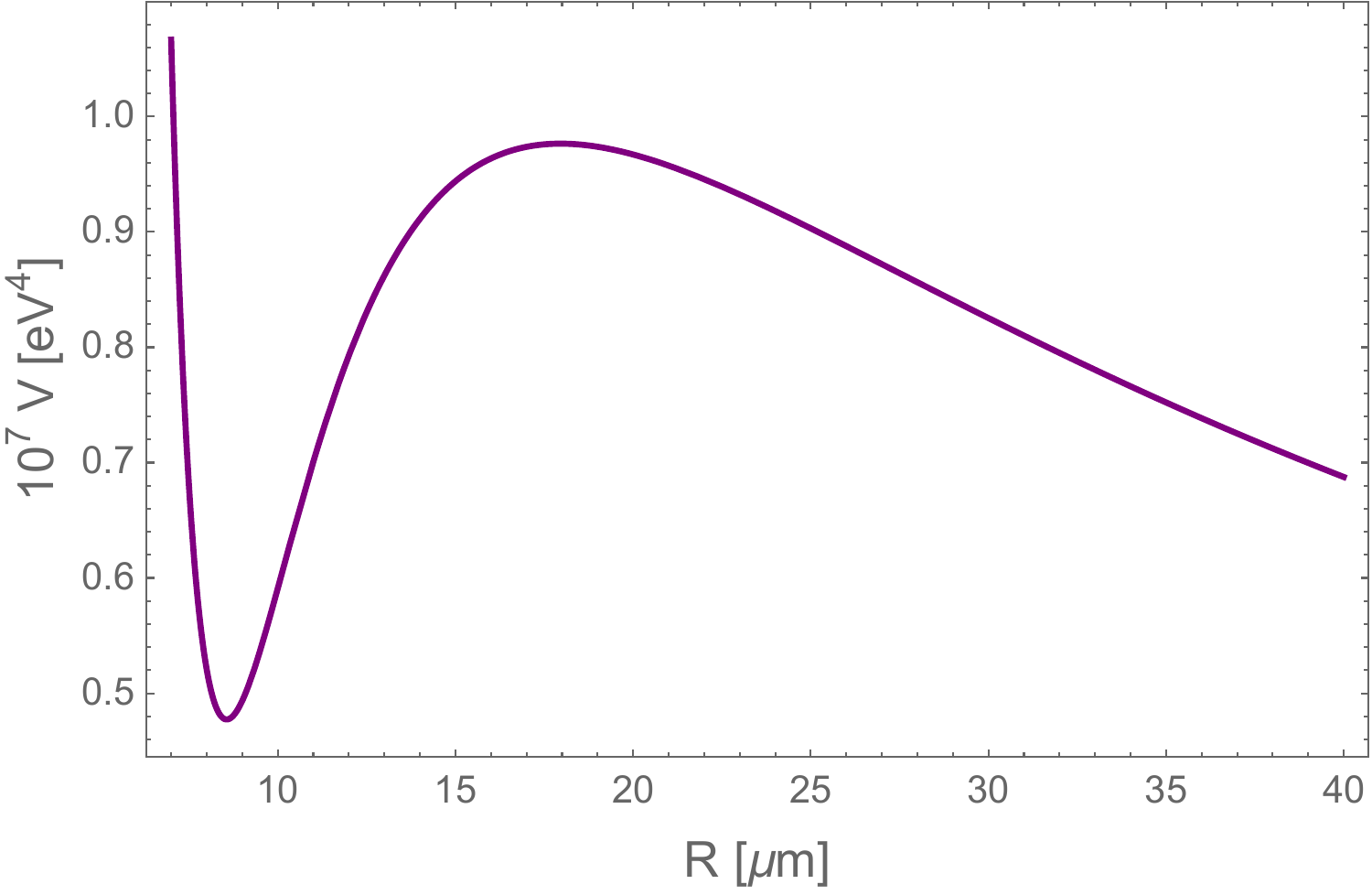}{0.8}
  \end{minipage}
\caption{The potential $V(R)$ for $(\Lambda_5^{\rm min})^{1/5} = 25~{\rm meV}$, $|T_4|^{1/4} =
  27~{\rm meV}$, and $N_F - N_B = 7$ (left), and $(\Lambda_5^{\rm min})^{1/5} = 25~{\rm meV}$, $|T_4|^{1/4} =
26~{\rm meV}$, and $N_F - N_B = 3$   (right). \label{fig:2}}
\end{figure}

In order to obtain a minimum with positive energy and a radius around the micron, the three terms of the potential should be comparable in magnitude.
In Fig.~\ref{fig:2} we show two  illustrative examples. The shape of
the potential of the left panel derives from $(\Lambda_5^{\rm min})^{1/5} = 25~{\rm meV}$, $|T_4|^{1/4} =
27~{\rm meV}$, and $N_F - N_B = 7$, where we have considered contributions
to the Casimir energy of the massless 5D graviton ($N_B = 5$) and three
neutrino states ($N_F = 12$). We have adopted Dirac bulk mass terms $\mu_i \alt R_{\rm max}^{-1}$ satisfying $ \Delta m_{21} < \Delta m_{32} \alt \mu_i$, where $\Delta m_{21}$ and $\Delta m_{32}$  are the mass differences required to explain the various neutrino oscillation phenomena, and $i = 1,2,3$~\cite{Lukas:2000wn,Lukas:2000rg,Carena:2017qhd}. As shown in~\cite{Anchordoqui:2023wkm}, this particular region of the parameter space allows us to avoid restrictive bounds from neutrino  oscillation experiments, which for $\mu_i \sim 0$,  constrain the size of the compact dimension $R \alt 0.2~\mu{\rm m}$~\cite{Machado:2011jt,Forero:2022skg}. The shape of the potential on the right panel corresponds to $(\Lambda_5^{\rm min})^{1/5} = 25~{\rm meV}$, $|T_4|^{1/4} =
26~{\rm meV}$, and $N_F - N_B = 3$, where we have considered the
lightest neutrino to be a massless Majorana brane state and so only two
neutrinos propagate in the bulk. 

In terms of the normalised field $\phi$ the potential (\ref{potential}) can be recast as
  \begin{equation}
    V(\phi) = 2 \pi \Lambda_5^{\rm min} r e^{- \sqrt {2/3} \phi} + T_4
    e^{-2 \sqrt{2/3} \phi} + \frac{1}{16 \pi^6} \frac{1}{r^4} (N_F - N_B)
    e^{-6\sqrt{2/3}\phi} \, .
  \end{equation}
It follows that the radion mass, $m_\phi \sim \sqrt{V''(0)}/M_p$, is  $m_\phi \sim 10^{-30}~{\rm eV}$ for the illustrative example in which $R \sim
  10~\mu{\rm m}$ and $m_\phi \sim 10^{-31}$ eV for $R \sim 8~\mu {\rm m}$. In addition, by comparing the maximum and minimum of the potential in
Fig.~\ref{fig:2}  we see that the oscillations of the radion around the minimum would have
a negligible contribution to the 4D cosmological constant $\Lambda_4$.

On the other hand, such a light scalar should have suppressed interactions to ordinary matter in order to avoid experimental limits on extra forces. In fact, in the absence of scalar potential, the radion from 5D is equivalent to a Brans-Dicke scalar with a parameter $\omega=-4/3$ that couples to the trace of the matter energy momentum tensor with a strength comparable to gravity. A way to suppress its coupling is by adding a 4D localised correction to the radion kinetic term\footnote{More general corrections to the radion effective action with the same goal were discussed in~\cite{Albrecht:2001cp}.}, in analogy with the brane contribution to the potential which is constant in the 5D frame 
given by the tension $T_4$:
\be
\delta S_{\rm radion}^{\rm localised}=\int [d^4x]\left\{ -\zeta 
\left({\partial R\over R}\right)^2-T_4\right\}\,,
\ee
with $\zeta$ a positive dimensionful parameter. A possible origin of $\zeta$ is through an expectation value of a brane field coupled to the radion kinetic terms via for instance the internal components of the 5D Ricci tensor. 
When added to the 5D action \eqref{Shd} (for $d=1$) dimensionally reduced to 4D, it modifies the Brans-Dicke parameter by a $1/R$ correction to $\omega=-4/3+\zeta/R$ (in $M_*$ units) that should be large enough to suppress the physical coupling of the radion to matter (upon appropriate rescaling that normalises its kinetic term).
Note that the modification of the radion kinetic term does not change the minimisation of the potential but lowers its mass by the same factor $\sim\sqrt\omega$ that suppresses its coupling; for this, one needs $\omega\simgt 10^3-10^4$.

In the absence of the $T_4$ contribution, the potential can still develop a minimum based on the negative contributions to the Casimir energy of the bosonic degrees of freedom, that produces a maximum at $R^{-1}_{\rm max}\sim (\Lambda_5^{\rm min})^{1/5}$, and the positive contribution of 5D fermions~\cite{Arkani-Hamed:2007ryu} (such as the bulk R-handed neutrinos) with appropriate bulk mass around the same scale; $(\Lambda_5^{\rm min})^{1/5}$ should then be around the eV scale. For example, if we only consider the massless 5D graviton $(N_B=5)$ and take  $(\Lambda_5^{\rm min})^{1/5} \sim 25~{\rm meV}$, the maximum appears at $R\sim 10~\mu{\rm m}$.

\section{Conclusions}

We have investigated the possibility that compact extra dimensions can obtain large size by higher dimensional inflation, relating the weakness of the actual gravitational force to the size of the observable universe. The proposed 5D inflationary set-up corresponds to a de Sitter (or approximate) solution of 5D Einstein equations, with cosmological constant and a 5D Planck scale $M_* \sim 10^9~{\rm GeV}$~\cite{Anchordoqui:2022svl}. All dimensions (compact and non-compact) expand exponentially in terms of the 5D proper time. The proposed set-up requires about $40$ e-folds to expand the 5th dimension from the fundamental length to the micron size. At the end of 5D inflation, or at any given moment, one can interpret the solution in terms of 4D fields using 4D Planck units from the relation $M_p^2=2\pi RM_*^3$, which amounts going to the 4D Einstein frame. This implies that if $R$ expands $N$ e-folds, then the 3D space would expand $3N/2$ e-folds as a result of a uniform 5D inflation. Altogether, the 3D space has expanded by about 60 e-folds to solve the horizon problem, while connecting this particular solution to the generation of large size extra dimension.

From the 4D perspective, the modulus potential is not constant and there are two fields: the 5D inflaton and the radion modulus. When inflation starts, the compactification radius is small and the 4D Planck mass is of order the 5D Planck scale (which one can consider to be the string scale), but when inflation ends, the radius is large and the 4D Planck scale arrives at the value observed today, $M_p \sim 10^{18}~{\rm GeV}$. Actually, from the 4D point of view, the 4D metric and the radion modulus evolve differently. In terms of the 4D proper time the scale factor of the 3D space expands as $t^3$ which corresponds to power law inflation, whereas the radion evolves logarithmically so that the size of the extra dimension expands as $t^2$.

We have shown that this solution gives a viable 4D cosmology consistent with CMB observations. This is because for any spacetime dimensionality, the 2-point function of a massless minimally coupled scalar field (such as a slow-rolling inflaton) in de Sitter space behaves logarithmically at distances larger than the cosmological horizon~\cite{Ratra:1984yq}, yielding a scale-invariant Harrison-Zel'dovich spectrum for distances smaller than the compactification length. Correspondingly, precision of CMB power spectrum applies to angles of less than about 10 degrees in the sky or equivalently to a Mpc scale distance. Now, the number of e-folds during radiation dominance is about 41  when the inflation scale is near (but below) the higher-dimensional Planck scale, say $M_I/{\rm GeV}\sim 10^7$ to $10^8$. This corresponds to an expansion of the 3D space by a factor of $10^{17}-10^{18}$, and thereby the critical CMB Mpc distance is shorten to $100-10$~km near the end of inflation. When this distance 
is converted to 5D units it is scaled down by a factor $(R_0/R)^{1/2}\sim 3\times 10^{-10}$ to $30 - 3~\mu$m, which is stunningly of order of the size of the 5th dimension at the end of inflation. We have shown that the predicted small-angle ($< 18^\circ$) CMB power spectrum is compatible with observations. Such an angle corresponds to a distance $\sim 4.5~{\rm Mpc}$ and multipole moment $\ell \simeq 10$; see Fig.~\ref{fig:1} for details. For smaller $\ell$ multipoles (larger angles), one obtains more power spectrum than standard 4D inflation, corresponding to a nearly vanishing spectral index, that the present data cannot distinguish due to large errors.  One caveat here is that the power spectrum is cosmic variance limited~\cite{Larson:2010gs}. However, even though cosmic variance prevents identification at low $\ell$, the transition region near $\pi kR_0$  could provide a signal for experiments in the near future. To determine $\ell$ multipoles in the transition region, an exhaustive transfer-function analysis (numerically solving the linearized Einstein-Boltzmann equations) would be required. 

One may wonder if the existence of the extra scalar from the 4D point of view can generate additional effects, such as iso-curvature perturbations and non-gaussianities. Since the radion modulus is however part of the 5D graviton, we expect these effects to be of the same order as the primordial gravity waves and thus to be suppressed. Although the extra scalar has an amplitude that is suppressed by the slow roll parameter, it contributes with its own power spectrum, destroying the simple power law of scalar perturbatins. In other words, the scalar perturbations become a sum of two contributions, one coming from the inflaton (that is determined by the slow roll parameters of the potential)  and another that comes from the radion  (that in principle has a different spectrum). Note that these two contributions have different momentum dependence so they cannot be parametrized as a simple power law, i.e. there will be a correction to the slow roll parameter that destroys the simple spectral index. Indeed, the spectral index will be a sum of two powers, with the second power being  suppressed by the slow roll parameter $\epsilon$. An expansion in $\epsilon$ can be parametrized as a $k$ variation of the spectal index, which is $\epsilon$ dependent. The power spectrum is then given by (\ref{P1}) and (\ref{P2}), up to a spectral tilt correction. Therefore, even if the gravitational waves are small and difficult to detect, precision  measurements of the scalar perturbations could indicate that there is a correction. A dedicated analysis is under investigation for confirmation and computation of the power spectrum of the tensor modes. After the end of 5D inflation, the radion has a runaway exponential potential quintessence-like with an exponent near the upper limit of the allowed value.

Alternatively, we have proposed a mechanism to stop the time evolution of the radion at the end of inflation, which is based on some general additional contributions to the scalar potential. The resulting minimisation is controlled by one parameter which is the magnitude of the contribution due to bulk field gradients, denoted $K$. For pedagogical reasons, we studied two extreme cases: \\
- When $K$ is of order the 5D fundamental scale, the radion mass reaches an upper bound proportional to the compactification scale around the eV region, while the vacuum energy can be tuned to an infinitesimal value as a result of cancellation among large contributions at the (sub)TeV scale.\\
- When $K$ vanishes, we have demonstrated that the radion can be stabilised in a local (metastable) dS vacuum by taking into account the contribution to the scalar potential of the Casimir energy, in the spirit of~\cite{Arkani-Hamed:2007ryu}. In general, one has three possible contributions. The first stems from the 5D cosmological constant $\Lambda_5^{\rm min}$ at the end of inflation, the second contribution comes from localised 3-branes/orientifolds with total tension $T_4$, and the last contribution corresponds to the Casimir energy from all 5D states. For large $R$, the contribution from the Casimir energy can be neglected,  and in the presence of a negative tension the potential develops a maximum at $R_{\rm max} = - T_4/(\pi \Lambda_5^{\rm min})$.\footnote{In the absence of tension, a maximum is still present due to the 5D graviton contribution to the Casimir potential.} We have shown that a minimum near the maximum can be accommodated using the Casimir energy contribution of light fermionic degrees of freedom driven by R-handed neutrino states with Dirac bulk masses $\alt 1/R_{\rm max}$. Thus, in this case a tuning of the vacuum energy is avoided but $\Lambda_5^{\rm min}$ and $T_4$ should be of order the eV scale.

Within the second set up however, the radion mass is estimated to be ${\cal O} (10^{-30}~{\rm eV})$. We have shown that the oscillations of the radion around the minimum would have a negligible contribution to the present dark energy $\Lambda_4$; for details see Fig.~\ref{fig:2}. On the other hand, consistency with the experimental limits on extra forces requires suppressing its coupling to matter. We have shown that such a suppression can arise from a 4D localised correction to the radion kinetic term with a large coefficient of order several thousands. Alternatively, it has been argued that an appropriate modification of such theories due to bulk quantum corrections can lead to a logarithmic scale (time) dependence of the Brans-Dicke parameter $\omega$ that suppresses the radion coupling to matter~\cite{Albrecht:2001cp}. 
For $K$ non-vanishing but less than the 4D fundamental scale, the radion mass varies between the two extreme cases and suppression of its coupling becomes less severe to not be necessary.

\section*{Acknowledgments}

We are extremely grateful to Nima Arkani-Hamed for enlightening discussions and collaboration during several stages of this work.\\
L.A.A. is supported by the U.S. National Science Foundation (NSF) Grant PHY-2112527. I.A. gratefully acknowledges support from the Simons Center for Geometry and Physics, Stony Brook University at which a last part of the research for this paper was performed.

\section*{Appendix}

The action of a massive scalar field $\Phi$ in a $D$-dimensional flat space is given by,
\begin{equation}
S_D = \int [d^Dx] \left[-\frac{1}{2} (\partial_M \Phi)^2 - m^2 \Phi^2 \right] \,,
\label{appendix-action}
\end{equation}  
where $m$ is the field's mass and $M$ runs from 0 to $(D-1)$. Introducing the dimensionless field $\chi = a^{D/2 -1} \ \Phi$, it follows that
\begin{equation}
  \dot \Phi  =  \left(1-\frac{D}{2} \right) \ a^{-D/2} \ \dot a \ \chi +  a^{1-D/2} \ \dot \chi  = a^{1-D/2} \left[\dot \chi + \left(1- \frac{D}{2} \right) \ \frac{\dot{a}}{a} \ \chi \right] \,,
\label{ap2}
\end{equation}
and
\begin{equation}
 \frac{\dot a}{a} \ \chi \dot \chi = \frac{1}{2} \ \chi^2 \ {\partial\over\partial\tau}{\left(\frac{a^{\!\!\!\!\sbt}}{a} \right)} \, .
\label{ap3}
\end{equation}    
Using (\ref{ap2}) and (\ref{ap3}) we can rewrite (\ref{appendix-action}) in Fourier
space as
\begin{equation}
  S = \int d\tau\, d^{^{D-1}}k\, \frac{1}{2} \left\{\dot{\tilde{\chi}}^2 + \left[\left(1 - \frac{D}{2}\right)^2 \left(\frac{\dot a}{a}\right)^2 + \left(1-\frac{D}{2}\right)  \frac{\partial}{\partial\tau}  \dot{\left(\frac{a^{\!\!\!\!\sbt}}{a} \right)} + k^2 - m^2 a^2 \right] \tilde{\chi}^2 \right\}\, .
  \label{ap-action2}
\end{equation}
For a power law dependence $a(\tau) \sim \tau^{-\alpha}$ with $\alpha>1$, corresponding to $a(t)\sim t^{\alpha\over\alpha-1}$, it follows that $\dot a/a \sim -\alpha/\tau$ and $\partial/\partial \tau(\dot a/a) \sim \alpha/\tau^2$, and then (\ref{ap-action2}) can be rewritten as
\begin{equation}
   S = \int d\tau\, d^{^{D-1}}k\, \frac{1}{2} \left\{\dot{\tilde{\chi}}^2 + \left[\alpha \left(1 - \frac{D}{2} \right) \left[\alpha \left(1 - \frac{D}{2}\right) + 1\right] \frac{1}{\tau^2} + k^2 - m^2 a^2 \right]\chi^2 \right\} \, .
\label{ap-action3}
 \end{equation}
 Using (\ref{ap-action3}) it is straightforward to check that for a massless scalar field the coefficient 15/4 is replaced by $L=\alpha(D/2-1)[\alpha(D/2-1)+1]=\alpha(\alpha+1)$ for $D=4$.
 
Duplicating the procedure for a $D$-dimensional dS space the action is found to be 
\begin{equation}
  S_{{\rm dS}_D} = \int d\tau \ d^{^{D-1}}k \ \frac{1}{2} \left\{\dot{\tilde{\chi}}^2 + \left[\left(\frac{D (D-2)}{4} - \frac{m^2}{ H^2}\right){1\over\tau^2} - k^2 \right] \tilde{\chi}^2 \right\}\, ,
\end{equation}
where $H = \dot a /a$~\cite{Ratra:1984yq}. In the case of a massive field  in dS$_D$, the coefficient 15/4 becomes $L=D(D-2)/4-m^2/H^2$. 

Following~\cite{Ratra:1984yq}, one finds that the order of the Bessel functions determining the correlation function~\eqref{Phi2} is given by $\nu=\sqrt{L+1/4}$, leading to a spectral index of the power spectrum $n_s=4-2\nu$.


\begin{thebibliography}{99}

\bibitem{Arkani-Hamed:1998jmv}
N.~Arkani-Hamed, S.~Dimopoulos and G.~R.~Dvali,
{\color{rossoCP3} The Hierarchy problem and new dimensions at a millimeter},
Phys. Lett. B \textbf{429} (1998), 263-272
doi:10.1016/S0370-2693(98)00466-3
[arXiv:hep-ph/9803315 [hep-ph]].

\bibitem{Antoniadis:1998ig}
I.~Antoniadis, N.~Arkani-Hamed, S.~Dimopoulos and G.~R.~Dvali,
{\color{rossoCP3} New dimensions at a millimeter to a Fermi and superstrings at a TeV},
Phys. Lett. B \textbf{436} (1998), 257-263
doi:10.1016/S0370-2693(98)00860-0
[arXiv:hep-ph/9804398 [hep-ph]].
 
\bibitem{Antoniadis:1988jn}
I.~Antoniadis, C.~Bachas, D.~C.~Lewellen and T.~N.~Tomaras,
{\color{rossoCP3} On Supersymmetry Breaking in Superstrings},
Phys. Lett. B \textbf{207} (1988), 441-446
doi:10.1016/0370-2693(88)90679-X;\\
I.~Antoniadis,
{\color{rossoCP3} A Possible new dimension at a few TeV},
Phys. Lett. B \textbf{246} (1990), 377-384
doi:10.1016/0370-2693(90)90617-F
 
\bibitem{Vafa:2005ui}
C.~Vafa,
{\color{rossoCP3} The String Landscape and the Swampland},''
[arXiv:hep-th/0509212 [hep-th]].

\bibitem{Ooguri:2006in}
H.~Ooguri and C.~Vafa,
{\color{rossoCP3} On the Geometry of the String Landscape and the Swampland},
Nucl. Phys. B \textbf{766}, 21-33 (2007)
doi:10.1016/j.nuclphysb.2006.10.033
[arXiv:hep-th/0605264 [hep-th]].

\bibitem{Dvali:2007hz}
G.~Dvali,
{\color{rossoCP3} Black holes and large $N$ species solution to the hierarchy problem},
Fortsch. Phys. \textbf{58}, 528-536 (2010)
doi:10.1002/prop.201000009
[arXiv:0706.2050 [hep-th]];\\
G.~Dvali and M.~Redi,
{\color{rossoCP3} Black hole bound on the number of species and quantum gravity at LHC},
Phys. Rev. D \textbf{77}, 045027 (2008)
doi:10.1103/PhysRevD.77.045027
[arXiv:0710.4344 [hep-th]].


\bibitem{Anchordoqui:2022svl}
L.~A.~Anchordoqui, I.~Antoniadis and D.~L\"ust,
{\color{rossoCP3} Aspects of the dark dimension in cosmology},
Phys. Rev. D \textbf{107} (2023) no.8, 083530
doi:10.1103/PhysRevD.107.083530
[arXiv:2212.08527 [hep-ph]].

\bibitem{ParticleDataGroup:2022pth}
R.~L.~Workman \textit{et al.} [Particle Data Group],
{\color{rossoCP3}  Review of Particle Physics},
PTEP \textbf{2022}, 083C01 (2022)
doi:10.1093/ptep/ptac097



\bibitem{Ratra:1984yq}
B.~Ratra,
{\color{rossoCP3} Restoration of Spontaneously Broken Continuous Symmetries in de Sitter Space-Time},
Phys. Rev. D \textbf{31} (1985), 1931-1955
doi:10.1103/PhysRevD.31.1931

\bibitem{Planck:2018vyg}
N.~Aghanim \textit{et al.} [Planck],
{\color{rossoCP3} Planck 2018 results VI: Cosmological parameters},
Astron. Astrophys. \textbf{641}, A6 (2020)
[erratum: Astron. Astrophys. \textbf{652}, C4 (2021)]
doi:10.1051/0004-6361/201833910
[arXiv:1807.06209 [astro-ph.CO]].






\bibitem{Harrison:1969fb}
E.~R.~Harrison,
{\color{rossoCP3} Fluctuations at the threshold of classical cosmology},
Phys. Rev. D \textbf{1}, 2726-2730 (1970)
doi:10.1103/PhysRevD.1.2726

\bibitem{Zeldovich:1972zz}
Y.~B.~Zeldovich,
{\color{rossoCP3} A Hypothesis, unifying the structure and the entropy of the universe},
Mon. Not. Roy. Astron. Soc. \textbf{160}, 1P-3P (1972)
doi:10.1093/mnras/160.1.1P

\bibitem{Montero:2022prj}
M.~Montero, C.~Vafa and I.~Valenzuela,
{\color{rossoCP3} The dark dimension and the Swampland},
JHEP \textbf{02} (2023), 022
doi:10.1007/JHEP02(2023)022
[arXiv:2205.12293 [hep-th]].

\bibitem{Dienes:1998sb}
K.~R.~Dienes, E.~Dudas and T.~Gherghetta,
{\color{rossoCP3}  Neutrino oscillations without neutrino masses or heavy mass scales: A Higher dimensional seesaw mechanism},
Nucl. Phys. B \textbf{557}, 25 (1999)
doi:10.1016/S0550-3213(99)00377-6
[arXiv:hep-ph/9811428 [hep-ph]].

\bibitem{Arkani-Hamed:1998wuz}
N.~Arkani-Hamed, S.~Dimopoulos, G.~R.~Dvali and J.~March-Russell,
{\color{rossoCP3}  Neutrino masses from large extra dimensions},
Phys. Rev. D \textbf{65}, 024032 (2001)
doi:10.1103/PhysRevD.65.024032
[arXiv:hep-ph/9811448 [hep-ph]].

\bibitem{Dvali:1999cn}
G.~R.~Dvali and A.~Y.~Smirnov,
{\color{rossoCP3}  Probing large extra dimensions with neutrinos},
Nucl. Phys. B \textbf{563}, 63-81 (1999)
doi:10.1016/S0550-3213(99)00574-X
[arXiv:hep-ph/9904211 [hep-ph]].


\bibitem{Arkani-Hamed:1999lsd}
N.~Arkani-Hamed, L.~J.~Hall, D.~Tucker-Smith and N.~Weiner,
{\color{rossoCP3}  Solving the hierarchy problem with exponentially large dimensions},
Phys. Rev. D \textbf{62} (2000), 105002
doi:10.1103/PhysRevD.62.105002
[arXiv:hep-ph/9912453 [hep-ph]].


\bibitem{Arkani-Hamed:2007ryu}
N.~Arkani-Hamed, S.~Dubovsky, A.~Nicolis and G.~Villadoro,
{\color{rossoCP3} Quantum Horizons of the Standard Model Landscape},
JHEP \textbf{06} (2007), 078
doi:10.1088/1126-6708/2007/06/078
[arXiv:hep-th/0703067 [hep-th]].


\bibitem{Lukas:2000wn}
A.~Lukas, P.~Ramond, A.~Romanino and G.~G.~Ross,
{\color{rossoCP3} Solar neutrino oscillation from large extra dimensions},
Phys. Lett. B \textbf{495}, 136-146 (2000)
doi:10.1016/S0370-2693(00)01206-5
[arXiv:hep-ph/0008049 [hep-ph]].


\bibitem{Lukas:2000rg}
A.~Lukas, P.~Ramond, A.~Romanino and G.~G.~Ross,
{\color{rossoCP3}  Neutrino masses and mixing in brane world theories},
JHEP \textbf{04}, 010 (2001)
doi:10.1088/1126-6708/2001/04/010
[arXiv:hep-ph/0011295 [hep-ph]].



\bibitem{Carena:2017qhd}
M.~Carena, Y.~Y.~Li, C.~S.~Machado, P.~A.~N.~Machado and C.~E.~M.~Wagner,
{\color{rossoCP3} Neutrinos in Large Extra Dimensions and Short-Baseline $\nu_e$ Appearance},
Phys. Rev. D \textbf{96} (2017) no.9, 095014
doi:10.1103/PhysRevD.96.095014
[arXiv:1708.09548 [hep-ph]].

\bibitem{Anchordoqui:2023wkm}
L.~A.~Anchordoqui, I.~Antoniadis and J.~Cunat,
{\color{rossoCP3} The Dark Dimension and the Standard Model Landscape},
[arXiv:2306.16491 [hep-ph]].


\bibitem{Lee:2020zjt}
J.~G.~Lee, E.~G.~Adelberger, T.~S.~Cook, S.~M.~Fleischer and B.~R.~Heckel,
{\color{rossoCP3} New test of the gravitational $1/r^2$ law at separations down to 52 $\mu$m},
Phys. Rev. Lett. \textbf{124}, no.10, 101101 (2020)
doi:10.1103/PhysRevLett.124.101101
[arXiv:2002.11761 [hep-ex]].

\bibitem{Hannestad:2003yd}
S.~Hannestad and G.~G.~Raffelt,
{\color{rossoCP3} Supernova and neutron star limits on large extra dimensions reexamined},
Phys. Rev. D \textbf{67}, 125008 (2003)
[erratum: Phys. Rev. D \textbf{69}, 029901 (2004)]
doi:10.1103/PhysRevD.69.029901
[arXiv:hep-ph/0304029 [hep-ph]].
  




\bibitem{efolds}
P. Kreling,
{\color{rossoCP3} The number of e-foldings of the radiation dominated epoch and the effect of cosmic transitions}
Bachelor thesis, May 2017
https://www2.physik.uni-bielefeld.de/fileadmin/user\_upload/theory\_e6/Bachelor\_Theses/BachelorArbeit\_PascalKreling.pdf;\\
G.~German,
{\color{rossoCP3} Measuring the expansion of the universe},
[arXiv:2005.02278 [astro-ph.CO]].

\bibitem{Cook:2015vqa}
J.~L.~Cook, E.~Dimastrogiovanni, D.~A.~Easson and L.~M.~Krauss,
{\color{rossoCP3}  Reheating predictions in single field inflation},
JCAP \textbf{04}, 047 (2015)
doi:10.1088/1475-7516/2015/04/047
[arXiv:1502.04673 [astro-ph.CO]].

\bibitem{Planck:2013jfk}
P.~A.~R.~Ade \textit{et al.} [Planck],
{\color{rossoCP3} Planck 2013 results. XXII. Constraints on inflation},
Astron. Astrophys. \textbf{571}, A22 (2014)
doi:10.1051/0004-6361/201321569
[arXiv:1303.5082 [astro-ph.CO]];\\
Y.~Akrami \textit{et al.} [Planck],
{\color{rossoCP3} Planck 2018 results. X. Constraints on inflation},
Astron. Astrophys. \textbf{641}, A10 (2020)
doi:10.1051/0004-6361/201833887
[arXiv:1807.06211 [astro-ph.CO]].


\bibitem{Sachs:1967er}
R.~K.~Sachs and A.~M.~Wolfe,
{\color{rossoCP3} Perturbations of a cosmological model and angular variations of the microwave background},
Astrophys. J. \textbf{147}, 73-90 (1967)
doi:10.1007/s10714-007-0448-9


\bibitem{Antoniadis:2011ib}
I.~Antoniadis, P.~O.~Mazur and E.~Mottola,
{\color{rossoCP3} Conformal Invariance, Dark Energy, and CMB Non-Gaussianity},
JCAP \textbf{09}, 024 (2012)
doi:10.1088/1475-7516/2012/09/024
[arXiv:1103.4164 [gr-qc]].


\bibitem{Dodelson:2003ft}
S.~Dodelson and F.~Schmidt,
{\color{rossoCP3}  Modern Cosmology},
(Academic Press, Elsevier, 2021),
ISBN 978-0-12-815948-4

\bibitem{Dienes:2011ja}
K.~R.~Dienes and B.~Thomas,
{\color{rossoCP3} Dynamical Dark Matter: I. Theoretical Overview},
Phys. Rev. D \textbf{85}, 083523 (2012)
doi:10.1103/PhysRevD.85.083523
[arXiv:1106.4546 [hep-ph]].


\bibitem{Gonzalo:2022jac}
E.~Gonzalo, M.~Montero, G.~Obied and C.~Vafa,
{\color{rossoCP3} Dark Dimension Gravitons as Dark Matter},
[arXiv:2209.09249 [hep-ph]].



\bibitem{Law-Smith:2023czn}
J.~A.~P.~Law-Smith, G.~Obied, A.~Prabhu and C.~Vafa,
{\color{rossoCP3} Astrophysical Constraints on Decaying Dark Gravitons},
[arXiv:2307.11048 [hep-ph]].



\bibitem{Anchordoqui:2023tln}
L.~A.~Anchordoqui, I.~Antoniadis and D.~L\"ust,
{\color{rossoCP3} Fuzzy Dark Matter, the Dark Dimension, and the Pulsar Timing Array Signal},
[arXiv:2307.01100 [hep-ph]].

\bibitem{Higuchi:1986py}
A.~Higuchi,
{\color{rossoCP3} Forbidden Mass Range for Spin-2 Field Theory in De Sitter Space-time},
Nucl. Phys. B \textbf{282}, 397-436 (1987)
doi:10.1016/0550-3213(87)90691-2

\bibitem{Barreiro:1999zs}
T.~Barreiro, E.~J.~Copeland and N.~J.~Nunes,
{\color{rossoCP3} Quintessence arising from exponential potentials},
Phys. Rev. D \textbf{61} (2000), 127301
doi:10.1103/PhysRevD.61.127301
[arXiv:astro-ph/9910214 [astro-ph]].


\bibitem{Machado:2011jt}
P.~A.~N.~Machado, H.~Nunokawa and R.~Zukanovich Funchal,
{\color{rossoCP3}  Testing for Large Extra Dimensions with Neutrino Oscillations},
Phys. Rev. D \textbf{84}, 013003 (2011)
doi:10.1103/PhysRevD.84.013003
[arXiv:1101.0003 [hep-ph]].

\bibitem{Forero:2022skg}
D.~V.~Forero, C.~Giunti, C.~A.~Ternes and O.~Tyagi,
{\color{rossoCP3}  Large extra dimensions and neutrino experiments},
Phys. Rev. D \textbf{106}, no.3, 035027 (2022)
doi:10.1103/PhysRevD.106.035027
[arXiv:2207.02790 [hep-ph]].




\bibitem{Albrecht:2001cp}
A.~Albrecht, C.~P.~Burgess, F.~Ravndal and C.~Skordis,
{\color{rossoCP3}  Exponentially large extra dimensions},
Phys. Rev. D \textbf{65} (2002), 123506
doi:10.1103/PhysRevD.65.123506
[arXiv:hep-th/0105261 [hep-th]];\\
{\color{rossoCP3}  Natural quintessence and large extra dimensions},
Phys. Rev. D \textbf{65} (2002), 123507
doi:10.1103/PhysRevD.65.123507
[arXiv:astro-ph/0107573 [astro-ph]].

\bibitem{Larson:2010gs}
D.~Larson, J.~Dunkley, G.~Hinshaw, E.~Komatsu, M.~R.~Nolta, C.~L.~Bennett, B.~Gold, M.~Halpern, R.~S.~Hill and N.~Jarosik, \textit{et al.}
{\color{rossoCP3}   Seven-year Wilkinson Microwave Anisotropy Probe (WMAP) observations: Power spectra and WMAP-derived parameters},
Astrophys. J. Suppl. \textbf{192} (2011), 16
doi:10.1088/0067-0049/192/2/16
[arXiv:1001.4635 [astro-ph.CO]].



\end{thebibliography}
\end{document}